\documentclass[12pt]{iopart}
\usepackage{iopams}
\usepackage{amsfonts}
\usepackage{amssymb}
\usepackage{setstack}

\eqnobysec

\begin{document}
%
%

\renewcommand{\a}{{\boldsymbol a}}
\renewcommand{\b}{{\boldsymbol b}}
\renewcommand{\c}{{\boldsymbol c}}
\renewcommand{\i}{{\boldsymbol i}}

\newcommand{\0}{{\boldsymbol 0}}
\newcommand{\1}{{\boldsymbol 1}}
\newcommand{\2}{{\boldsymbol 2}}
\newcommand{\3}{{\boldsymbol 3}}

\newcommand{\bj}{{\boldsymbol j}}

\newcommand{\ba}{{\boldsymbol{\alpha}}}
\newcommand{\bb}{{\boldsymbol \beta}}
\newcommand{\bc}{{\boldsymbol \gamma}}
\newcommand{\bd}{{\boldsymbol \delta}}

\newcommand{\A}{{\boldsymbol A}}
\newcommand{\B}{{\boldsymbol B}}
\newcommand{\C}{{\boldsymbol C}}
\newcommand{\D}{{\boldsymbol D}}
\newcommand{\E}{{\boldsymbol E}}
\newcommand{\F}{{\boldsymbol F}}


\newcommand{\tlam}{\tilde{\lambda}}
\newcommand{\m}[1]{\mathcal{#1}}
\newcommand{\ty}{\tilde{y}}
\newcommand{\bigs}{\bigskip}
\newcommand{\var}[2]{\frac{\delta{#1}}{\delta{#2}}}
\newcommand{\ap}{\approx}
\newcommand{\Si}[3]{\tilde{\Sigma}_{#1}^{\,\,\,#2#3}}
\newcommand{\ud}{\,\mathrm{d}}
\newcommand{\half}{\textstyle{\frac{1}{2}}}
\newcommand{\pa}{{\phantom{a}}}
\newcommand{\pd}[1]{
        {\frac{\partial}{\partial{#1}}}}


\title[2+2 Ashtekar variables]{Hamiltonian analysis of the double null 2+2 decomposition of Ashtekar variables}
\author{R A d'Inverno, P Lambert   and J A Vickers}
\address{School of Mathematics, University of Southampton,
Southampton, SO17~1BJ, UK}
\ead{J.A.Vickers@maths.soton.ac.uk}
\begin{abstract}
We derive a canonical analysis of a double null 2+2 Hamiltonian description
of General Relativity in terms of complex self-dual 2-forms and the associated
$SO(3)$ connection variables. The algebra of first class constraints is
obtained and forms a Lie algebra that consists of two constraints that
generate diffeomorphisms in the two surface, a constraint that generates
diffeomorphisms along the null generators and a constraint that generates
self-dual spin and boost transformations.
\end{abstract}
\submitted{\CQG}
\pacs{0420Fy}
\maketitle
%

\section{Introduction}
Early attempts at using the methods of canonical quantisation to provide a
theory of quantum gravity were based on the ADM formalism \cite{adm:60},
however it was soon realised that the non-polynomial nature of the
constraints prevented one from moving from the classical canonical analysis
to a quantum description. A significant advance was made by Ashtekar 
\cite{aa:87}
with the introduction of a new formulation of General Relativity in which
the variables are taken to be the components of a complex $SO(3)$
connection with conjugate momenta the components of a triad of vector
densities. This had the effect of presenting General Relativity in a form
similar to Yang-Mills theory and also making the constraints polynomial in
the variables. However despite substantial progress in developing a quantum
theory of gravity based on this approach \cite{aa2} there are still some
difficulties that arise from the Hamiltonian constraint and the fact that
the first class constraint algebra is not a Lie algebra.
(Although see the work of Thiemann \cite{tt} for recent 
progress with this problem).

It was pointed out by Goldberg {\it et al} \cite{jg&dr:92} that if one
works with a null foliation the Hamiltonian constraint becomes second class
and that the algebra of first class constraints then becomes a Lie
algebra. However if one chooses a null 3+1 decomposition there is no
natural projection operator associated with the foliation. A further
problem is that the transformations which preserve the foliation is the
group of null rotations which is algebraically awkward to work with. Both
these problems are avoided if one works with a double null 2+2 description
in which the projection is well defined and the relevant group is that of
spin and boost transformations whose group structure is simply
multiplication of non-zero complex numbers. An additional reason for
working with a 2+2 formalism is that an analysis of the field equations
shows that the gravitational degrees of freedom may be chosen to lie in the
conformal structure of the induced 2-metric \cite{rd&js:80}.

In this paper we will derive  a Hamiltonian description of a double null
2+2 decomposition of General Relativity given in terms of Ashtekar's 
formulation using a complex $S0(3)$ connection. We then use the Dirac 
theory of constraints to construct the first class algebra and give a
geometric interpretation to the four first class constraints. The
formulation will be based on an earlier paper by d'Inverno and Vickers 
\cite{rd&jv:95} (which we will refer to as paper I) 
in which we presented a double null 2+2 Lagrangian
description of General Relativity using a version of Ashtekar variables
based on self-dual 2-forms and the corresponding first order action given
by Jacobson and Smolin \cite{tj&ls:88} and also Samuel \cite{samuel}. 
Unlike some recent treatments using real connection variables we use 
the original Ashtekar formulation in which the manifold is real but 
we allow complex solutions of the field equations. However once the
canonical analysis has been carried out reality conditions are imposed to
limit the solutions to  real solutions of Einstein's equations.
In section 2 we review the basic
variables used in paper I and in the following section we briefly describe
the Lagrangian approach. The Hamiltonian is then introduced and in
subsequent sections we demonstrate that this gives us all the Einstein
equations as well as the structure equations for the $SO(3)$ connection.  
In section 7 we use the Dirac theory of constraints to obtain the first
class algebra and interpret the constraints geometrically.

\section{The 2+2 tetrad}
We start by briefly reviewing the notation and geometric variables
employed in \cite{rd&jv:95}.
Throughout the paper Greek indices run from 0 to 3, early Latin indices
$(a,b,\ldots)$ run from 0 to 1, middle Latin indices $(i,j,\ldots)$ run
from 2 to 3, uppercase Latin indices $(A,B,\ldots)$ run from 1 to 3
and tetrad indices will be written in bold.
Let M be a four-dimensional orientable manifold with metric
$g$ of signature $(+1,-1,-1,-1)$. 
A foliation of codimension two can be described by two closed 1-forms
$n^{\0}$ and $n^{\1}$. Thus locally \cite{rd&js:80}
\begin{equation}
dn^{\a}=0 \Longleftrightarrow n^{\a}=d\phi^{\a}.
\end{equation}
The two 1-forms generate hypersurfaces defined by
\numparts
\begin{eqnarray}
\{\Sigma_{\0}\}: \phi^{\0}(x^\alpha)&=\hbox{constant},\\
\{\Sigma_{\1}\}: \phi^{\1}(x^\alpha)&=\hbox{constant},
\end{eqnarray} \endnumparts
respectively. These hypersurfaces define a family of 2-surfaces $\{S\}$ by
\begin{equation}
\{S\}=\{\Sigma_{\0}\}\cap\{\Sigma_{\1}\}.
\end{equation}
We restrict attention to the case when $\{S\}$ is spacelike and
denote the family of two dimensional timelike spaces orthogonal to $\{S\}$
at each point by $\{T\}$. Let $n_{\a}$ be the dyad basis of vectors dual
to $n^{\a}$ in $\{T\}$, so that
\begin{equation}
n_{\a}^{\pa \alpha}n^{\b}_{\pa \alpha}=\delta_{\a}^{\b}.
\end{equation}
We define projection operators into $\{S\}$ and $\{T\}$ by
\numparts\begin{eqnarray}
B^{\alpha}_{\beta}&
=\delta^{\alpha}_{\beta}-n_{\a}^{\pa\alpha}n^{\a}_{\pa\beta}, \\
T^{\alpha}_{\beta}&=n_{\a}^{\pa\alpha}n^{\a}_{\pa\beta}.
\end{eqnarray} \endnumparts
The 2-metric induced on $\{S\}$ is given by the projection
\begin{equation}
\phantom{a}^2g_{\alpha\beta}
=B^{\gamma}_{\alpha}B^{\delta}_{\beta}g_{\gamma\delta}
=B_{\alpha\delta}B^{\delta}_{\beta}
=B_{\alpha\beta}
\end{equation}
and we use the 
$n_{\a}$ to define a $2\times 2$ matrix of scalars $N_{\a\b}$ by
\begin{equation}
N_{\a\b}=g_{\alpha\beta}n_{\a}^{\pa \alpha}n_{\b}^{\pa \beta}.
\end{equation}
The elements $N_{\0\0}$ and $N_{\1\1}$ define the lapses of
$\{S\}$ in $\{\Sigma_{\0}\}$ and $\{\Sigma_{\1}\}$, respectively.

We now choose a pair of vectors $E_{\a}$ which connect neighbouring
2-surfaces in $\{S\}$. We choose them such that
\begin{equation}
n^{\a}_{\pa\alpha}E_{\b}^{\pa\alpha}=\delta^{\a}_{\b},
\end{equation}
which defines $E_{\a}$ up to an arbitrary shift vector $b_{\a}$, i.e.
\begin{equation}
E_{\a}=n_{\a}+b_{\a}
\end{equation}
with
\begin{equation}
n^{\a}_{\pa\alpha}b_{\c}^{\pa\alpha}=0.
\end{equation}
In suitably adapted coordinates this results in the 2$+$2 
decomposition of the contravariant metric
\begin{equation}
g^{\alpha\beta}=
\left(\begin{array}{cc}
N^{ab}&-N^{ab}b_{\pa b}^{i}\\
-N^{ab}b_{\pa b}^{i}&\phantom{a}^2g^{ij}+N^{ab}b_{\pa a}^{i}b_{\pa b}^{j}\\
\end{array}\right).
\end{equation}

In order to give a 2+2 description of Ashtekar variables we start with
a 4-dimensional description of General Relativity in terms of
self-dual 2-forms and a complex $SO(3)$ connection (see
e.g. \cite{dg:94}) and then project this into $\{S\}$ and $\{T\}$.
Our starting point is a Newman-Penrose null tetrad $(e_{\ba})$ for the
metric with dual basis of 1-forms $(\theta^{\ba})$ so that
\begin{equation}
ds^2 = \eta_{\ba\bb}\theta^{\ba}\otimes\theta^{\bb}
\end{equation}
where
\begin{equation}
\eta_{\ba\bb}=
\left(\begin{array}{cccc}
0&1&0&0\\
1&0&0&0\\
0&0&0&-1\\
0&0&-1&0\\
\end{array}\right).
\end{equation}
(For real Lorentzian metrics $\theta^{\0}$ and $\theta^{\1}$ are real and 
$\theta^{\3}$ is the complex conjugate of $\theta^{\2}$.) 
This tetrad is then used to define a basis of self-dual 2-forms given by 
\begin{eqnarray}\begin{array}{ll}
    S^\1&=\half(\theta^{\1}\wedge\theta^{\0}+\theta^{\3}\wedge\theta^{\2}),\\
    S^\2&=\theta^{\1}\wedge\theta^{\2},\\
    S^\3&=\theta^{\3}\wedge \theta^{\0}.\\
\end{array}\label{2-forms}\end{eqnarray}
It was shown in paper I that a general basis of
1-forms with 2+2 decomposition is given by
\numparts
\begin{eqnarray}
    \theta^{\a}&=\mu^{\a}_{\pa b}dx^b
        +\alpha^{\a}_{\pa i} (dx^i+s^i_{\pa b}dx^b),\\
    \theta^{\i}&=\nu^{\i}_{\pa j}(dx^j+s^j_{\pa a}dx^a).
\end{eqnarray}
\label{2.2:basis}
\endnumparts 
The four $2\times 2$ matrix variables $\mu^{\a}_{\pa b}$, $\nu^{\i}_{\pa j}$,
$s^j_{\pa a}$ and $\alpha^{\a}_{\pa i}$ constitute the 16 degrees of
freedom corresponding to the 10 metric variables and the 6 Lorentz 
degrees of freedom. The dual basis is then given by
\numparts\begin{eqnarray}
    e_{\a}&=u_{\a}^{\pa b}\big( \pd{x^b}-s^i_{\pa b}\pd{x^i} \big),\\
    e_{\i}&=v_{\i}^{\pa j}\pd{x^j}+ \alpha^{\a}_{\pa j}v^j_{\pa \i}\left(
        u_{\a}^{\pa b}s^j_{\pa b}\pd{x^j}- u_{\a}^{\pa b}\pd{x^b}\right).
\end{eqnarray}\endnumparts
where the $2\times2$ matrices $u_{\a}^{\pa b}$ and $v_{\i}^{\pa
j}$ are defined to be inverses of $\mu^{\a}_{\pa b}$ and
$\nu^{\i}_{\pa j}$ respectively, so that
\begin{eqnarray}
    u_{\a}^{\pa b}\mu^{\a}_{\pa c}&=\delta^b_c, \qquad
        u_{\a}^{\pa c}\mu^{\b}_{\pa c}=\delta^{\b}_{\a},\\
    v_{\i}^{\pa j}\nu^{\i}_{\pa k}&=\delta^j_k, \qquad
        v_{\i}^{\pa k}\nu^{\bj}_{\pa k}=\delta^{\bj}_{\i}.
\end{eqnarray}

Considerable simplification can be obtained by working in an adapted
frame in which the $e_\i$ are tangent to $\{S\}$. With our choice of
frame this requires that the $\alpha$'s vanish, i.e.
\begin{equation}
\alpha^\ba_{\pa i}=0.
\end{equation}
This is not a restriction on the metric but purely on the choice of
frame and reduces the full six parameter group of Lorentz
transformations to the two parameter subgroup of spin and boost
transformations. It is worth pointing out that this is different from
the 3+1 null formulation of Goldberg {\it et al} \cite{jg&dr:92}
where choosing an
adapted frame also leads to coordinate conditions and so must be
imposed using a Lagrange multiplier if one is to obtain all the
Einstein equations. With our choice of an adapted frame 
$\mu^{\a}_{\pa b}$ and $s^j_{\pa a}$ generate the lapses and shifts while 
$\nu^{\i}_{\pa j}$ generates the 2-metric.

The next step is to impose the condition that both $x^0$ and $x^1$ are
null coordinates. In an adapted frame this is simply the condition
that
\numparts
\begin{eqnarray}
    g^{\0\0}&=g^{\alpha\beta}\theta^\0_{\pa \alpha}\theta^\0_{\pa \beta}
        =2\mu^\0_{\pa 0}\mu^\0_{\pa 1}=0,\\
    g^{\1\1}&=g^{\alpha\beta}\theta^\1_{\pa \alpha}\theta^\1_{\pa \beta}
        =2\mu^\1_{\pa 0}\mu^\1_{\pa 1}=0.
\end{eqnarray}
\endnumparts
The volume form is given by
\begin{equation}
    V=-i\theta^\0\wedge\theta^\1\wedge\theta^\2\wedge\theta^\3
     =-i\mu\nu dx^0\wedge dx^1\wedge dx^2\wedge dx^3,
\end{equation}
(where $\mu=\det(\mu^{\a}_{\pa b})$ and $\nu=\det(\nu^{\i}_{\pa j})$ )
which implies that $\mu$ and $\nu$ are non-zero. Therefore in order 
to satisfy the double
null slicing conditions as well as the condition that $\mu=\mu^\0_{\pa
0}\mu^\1_{\pa 1}-\mu^\0_{\pa 1}\mu^\1_{\pa 1}$ is non zero, we require
that either
\begin{equation}
    \mu^\0_{\pa0}=\mu^\1_{\pa 1}=0 \label{nonslicing}
\end{equation}
or
\begin{equation}
    \mu^\0_{\pa 1}=\mu^\1_{\pa 0}=0\label{2.2:slicing}
\end{equation}
are satisfied. Although we choose to require \eref{2.2:slicing},
there is no loss of generality because a change to the other
condition \eref{nonslicing} is equivalent to interchanging the 
coordinates $x^0$ and $x^1$.

We are now in a position to write down the self-dual 2-forms in terms
of the metric variables. These are given as follows
\numparts
\begin{eqnarray}
\fl S^\1&=\half(\theta^{\1}\wedge\theta^{\0}
        +\theta^{\3}\wedge\theta^{\2}) \nonumber \\
\fl    &=\half[(\mu^{\1}_{\pa a}\mu^{\0}_{\pa b} +\nu^{\3}_{\pa i}
        s^i_{\pa a}\nu^{\2}_{\pa j}s^j_{\pa b})dx^a\wedge dx^b
        -(\nu^{\2}_{\pa j}s^j_{\pa a}\nu^{\3}_{\pa i}
        -\nu^{\3}_{\pa j}s^j_{\pa a}\nu^{\2}_{\pa i})
        dx^a\wedge dx^i],\\
\fl    S^\2&=\theta^{\1}\wedge\theta^{\2} \nonumber \\
\fl       &=(\mu^{\1}_{\pa a}\nu^{\2}_{\pa i}s^i_{\pa b})
        dx^a\wedge dx^b
        +(\mu^{\1}_{\pa a}\nu^{\2}_{\pa i})dx^a\wedge dx^i
        +(\nu^{\3}_{\pa i}\nu^{\2}_{\pa j})dx^i\wedge dx^j,\\
\fl    S^\3&=\theta^{\3}\wedge \theta^{\0} \nonumber \\
\fl    &=\nu^{\3}_{\pa j}(dx^j+s^j_{\pa a}dx^a)
        \wedge(\mu^{\0}_{\pa b}dx^b) \nonumber \\
\fl    &=(\nu^{\3}_{\pa i}s^i_{\pa a}\mu^{\0}_{\pa b})dx^a\wedge dx^b
          -(\nu^{\3}_{\pa i}\mu^{\0}_{\pa a})dx^a\wedge dx^i.
\label{2.2:Stetrad}
\end{eqnarray}
\endnumparts
As is usual in the Ashtekar formalism we define a densitised version
of $S^\A$ by introducing the quantities
\begin{equation}
    \Si \A\alpha\beta=\half\epsilon^{\alpha\beta\gamma\delta}
        S^\B_{\pa \gamma\delta}g_{\A\B},
\end{equation}
where $g_{\A\B}$ is the $SO(3)$ invariant metric (defined by equation
(4.3a) in paper I). We may now express the sigma variables
in terms of the tetrad variables using~\eref{2.2:Stetrad}.

\section{The 2+2 Lagrangian}

In place of the usual Einstein-Hilbert action, we work with the
complex first-order action appropriate to self-dual 2-forms 
used by Jacobson and Smolin \cite{tj&ls:88}
\begin{equation}
    L=\int R^\A\wedge S^\B g_{\A\B},
\label{action}
\end{equation}
where $R^\A$ is the curvature 2-form of the $SO(3)$ connection $\Gamma^\A$.
The connection 1-forms $\Gamma^\A$ have 2+2 decomposition 
\begin{eqnarray}
    \Gamma^\A=\Gamma^\A_{\pa \mu}dx^{\mu}
        =A^\A_{\pa i}dx^i+B^\A_{\pa a}dx^a.
\label{2.2conn}\end{eqnarray}
The curvature 2-forms $R^\A$ are defined by
\begin{equation}
    R^\A=\ud\Gamma^\A+\eta^\A_{\pa \B\C}\Gamma^\B\wedge\Gamma^\C.
\end{equation}
When written in terms of the $SO(3)$ covariant derivative $D$ these 
have 2+2 decomposition
\numparts\begin{eqnarray}
    R^\A_{\pa ab}&=B^\A_{\pa b,a}-D_b B^\A_{\pa a},\\
    R^\A_{\pa ai}&=A^\A_{\pa i,a}-D_i B^\A_{\pa a},\\
    R^\A_{\pa ij}&=A^\A_{\pa j,i}-D_j A^\A_{\pa i}.
\end{eqnarray}\endnumparts
We may now write the $SO(3)$  action \eref{action} in terms of 
our variables as:
\begin{equation}
    L=\int (R^\A_{\pa 01}\Si\A01
        +R^\A_{\pa 23}\Si\A23 +R^\A_{\pa ai}\Si \A ai) \ud^4x.
\label{action2}
\end{equation}
Note that in an adapted frame
\begin{equation}
    (\Si \101,\Si \201, \Si \301) =(-\nu,0,0)\label{2.2Siteta}
\end{equation}
so that the term $R^\A_{\pa 01}\Si\A01$ simplifies to $\nu R^\1_{\pa 01}$.

In paper I we adopted a description similar to that of Goldberg
{\it et al} \cite{jg&dr:92} in their 3+1 null formulation and considered
the configuration space to be given in
terms of $\mu^{\a}_{\pa b}$ and $s^j_{\pa a}$ (the lapse and shift
parts of the frame) but replaced the $\nu^{\i}_{\pa j}$ variables 
by the mixed terms $\Si\A ai$ in the densitised 2-forms. 
As shown in paper I these variables are not
independent but must satisfy eight constraints which are given by
\begin{equation}
C^i\equiv\mu^{\0}_{\pa a}\Si \2ai=0,        \label{2.2:Ci}
\end{equation}    
\begin{equation}
    \tilde{C}^i\equiv\mu^{\1}_{\pa a}\Si \3ai=0, \label{2.2:tCi}
\end{equation}
and
\begin{equation}
C^i_{\pa a}\equiv s^i_{\pa a}\Si \101
        -\epsilon_{ab}\Si \1bi=0.\label{2.2:Cia}
\end{equation}
As is explained in \cite{rd&js:80} a further simplification in the 2+2
formalism is obtained by introducing the conformal factor $\nu$ of the
2-metric as an explicit variable. This results in one further constraint
\begin{equation}
    \hat{C}\equiv \Si \2ai\Si \3bj
        \epsilon_{ab}\epsilon_{ij} -\mu\nu=0.\label{2.2:hC}
\end{equation}

We are now in a position to write down the primary Lagrangian. It is
obtained from the action given above. The double null slicing
conditions are imposed through the Lagrange multipliers $\rho$ and
$\tilde \rho$, the adapted frame conditions, which require 
that $\Si 201$ and $\Si 301$ vanish, are imposed through 
Lagrange multipliers $\tau^2$ and $\tau^3$ 
while the constraints \eref{2.2:Ci}, \eref{2.2:tCi},
\eref{2.2:Cia} and  \eref{2.2:hC} are imposed through the corresponding
$\lambda$ Lagrange multipliers. The end result is
\begin{equation}
\begin{array}{ll}
\fl     L&=\int\left(\Si\A0i A^\A_{i,0}+\Si \A01 B^\A_{1,0}
        +B^\A_0 D_1 \Si\A01+ B^\A_0 D_i\Si\A0i
        -\mu R^\1_{\pa 23} -s^i_{\pa 0} R^\A_{\pa ij}\Si\A0j \right.\\
\fl    &-s^i_{\pa 1}(R^\2_{\pa ij}\Si\21j +R^\3_{\pa ij}\Si\31j)
        +R^\A_{\pa 1i}\Si\A1i +\lambda_iC^i +\tlam_i\tilde{C}^i
        +\hat{\lambda}\hat{C}+\lambda^a_i C^i_a \\
\fl    &\left.+\rho(\mu^0_{\pa 1})^2
        +\tilde{\rho}(\mu^1_{\pa 0})^2 +\tau^2\tilde\Sigma_\2^{\pa 01}
        +\tau^3\tilde\Sigma_\3^{\pa 01} \right)\ud^4x, \\
\end{array}
\end{equation}
where we have explicitly written out the curvature terms 
$R^\A_{\pa 01}$ and $R^\A_{\pa 0i}$ since they contain time derivatives 
of the connection.
It is worth noting that if one imposes the double null slicing
condition the constraints $C^i$ and $\tilde{C}^i$ simplify and become
\begin{equation}
    C^i=\Si \20i=0, \hspace{2cm} \tilde{C}^i=\Si \31i=0.
\end{equation}

In this formulation, the configuration space consists of the variables
$\nu$, $\mu^{\a}_{\pa b}$, $s^{\i}_{\pa a}$, $\tilde\Sigma_A^{\pa ai}$,
$A^A_{\pa i}$ and $B^A_{\pa a}$ which are required to satisfy a total of 13
constraints which are imposed through the Lagrange multipliers
$\hat{\lambda}$, $\lambda_i$, $\tilde{\lambda}_i$, $\lambda^a_{\pa i}$,
$\rho$, $\tilde{\rho}$ and $\tau^i$.  It was shown in paper I that
variation of this Lagrangian with respect to the variables $A^A_{\pa i}$
and $B^A_{\pa a}$ produces the structure equations of the $SO(3)$ connection
while variation with respect to the other variables allows one to eliminate
the Lagrange multipliers and obtain all the Einstein field equations. In
this paper we will give the corresponding Hamiltonian description and carry
out an analysis of the constraints to obtain a full canonical analysis of
the 2+2 Hamiltonian description in terms of Ashtekar type variables.

\section{Hamiltonian description}
The Lagrangian density is of the form $\m
L=p^\lambda\dot{q}_\lambda-\m H$, and therefore we can see directly
that the canonical variables are $A^\A_i$ and $B^\A_1$, and have the
respective momenta $\Si \A0i$ and $\Si \A01$. We can therefore simply read off
the Hamiltonian density which is given by

\begin{equation}
\begin{array}{ll}
\fl \m H &=\mu R^\1_{\pa 23}+s^i_{\pa 0} R^\A_{\pa ij}\Si\A0j
          +s^i_{\pa 1}(R^\2_{\pa ij}\Si\21j +R^\3_{\pa ij}\Si\31j)
          -R^\A_{\pa 1i}\Si\A1i \\
\fl      &-B^\A_0( D_1\Sigma_\A^{\pa \0\1}
          +D_i\Sigma_\A^{\pa 0i}) +\lambda_iC^i +\tlam_i\tilde{C}^i 
          +\hat{\lambda}\hat{C}+\lambda^a_i C^i_a \\
\fl      &+\rho(\mu^0_{\pa 1})^2
          +\tilde{\rho}(\mu^1_{\pa 0})^2
          +\tau^\2\Si\201 +\tau^\3\Si\301.
\end{array}
\end{equation}
The canonical Poisson brackets are then given by 
\numparts\begin{eqnarray}
    \left\{ A^\A_i(x),\Si\B0j(\ty)\right\}
        &=\delta^\A_\B\delta_i^j \delta(x,\ty)\\
    \left\{ B^\A_1(x),\Si\B01(\ty)\right\}&=\delta^\A_\B\delta(x,\ty).
\end{eqnarray}\endnumparts

In the Hamiltonian given above the variables $\mu^\a_{\pa b}, s^i_{\pa a}, 
\Si \A1i$ and $B^\A_0$ are cyclic variables. Because of the structure
equations and the Bianchi identities the equations obtained by variation
with respect to these variables are propagated by the primary Hamiltonian
and we do not need to include them in the full canonical analysis but can
consider them as if they were multipliers (as is done with the lapse and
shift in the standard ADM treatment). This `shortcut' procedure is
described in \S3.2 in the article on {\it Canonical Gravity} by Isenberg
and Nester \cite{gr&g}.  A similar observation was made by
Goldberg {\it et al} \cite{jg&dr:92} for the corresponding variables in the
null 3+1 case. It is also important to note that some of the constraints
introduced into the primary Hamiltonian are not constraints on the
canonical variables, but simply on the cyclic variables, and so 
may be treated as
multiplier equations. As a result of this we have a phase space which
consists of 18 variables $A^\A_i$, $B^\A_1$, $\Si\A0i$ and
$\Si\A01$. Furthermore only four of the original
thirteen constraints are actually primary constraints.
\begin{equation}
    C^i=0,\qquad \Si \201=0,\qquad \Si \301=0,\label{2.2:prim}
\end{equation}

We now start the constraint analysis algorithm by varying the
Hamiltonian with respect to the cyclic variables. This leads to
the equations

\numparts\begin{eqnarray}
    \var{H}{\mu^\0_{\pa 0}}&= -\mu^\1_{\pa 1}R^\1_{\pa 23}
        -\mu^\1_{\pa 1}\Si\101\hat{\lambda}
        -\lambda_i\Si\20i,\label{eq:dotp00}\\
    \var{H}{\mu^\1_{\pa 1}}&= -\mu^\0_{\pa 0}R^\1_{\pa 23}
        -\mu^\0_{\pa 0}\Si\101\hat{\lambda}
        -\tilde{\lambda}_i\Si\31i,\label{eq:dotp11}\\
    \var{H}{\mu^\0_{\pa 1}}&= \mu^\1_{\pa 0}R^\1_{\pa 23}
        +\mu^\1_{\pa 0}\Si\101\hat{\lambda}
        -\lambda_i\Si\21i, \label{eq:dotp10}\\
    \var{H}{\mu^\1_{\pa 0}}&= \mu^\0_{\pa 1}R^\1_{\pa 23}
        +\mu^\0_{\pa 1}\Si\101\hat{\lambda}
        -\tlam_i\Si\30i,\label{eq:dotp01}
\end{eqnarray}\endnumparts

\numparts\begin{eqnarray}
    \var{H}{s^i_{\pa 0}}&= R^\A_{\pa ij}\Si\A0j
        +\lambda^0_i\Si\101,\label{eq:dotk0p}\\
    \var{H}{s^i_{\pa 1}}&= R^\2_{\pa ij}\Si\21j
        +R^\3_{\pa ij}\Si\31j
        +\lambda^1_i\Si\101,\label{eq:dotk1p}
\end{eqnarray}\endnumparts

\numparts \begin{eqnarray}
    \var{H}{\Si\11p}&=R^\1_{\pa 1p}
        +\lambda^0_p,\label{eq:dotL11p}\\
    \var{H}{\Sigma_\2^{\pa 1p}}&=R^\2_{\pa 1p}
        -R^\2_{\pa jp}s^j_{\pa 1}-\lambda_p\mu^\0_{\pa 1}
        +\hat{\lambda}\Si\30j\epsilon_{pj},\label{eq:dotL21p}\\
    \var{H}{\Si\31p}&=R^\3_{\pa 1p}
        -R^\3_{\pa jp}s^j_{\pa 1}-\tilde{\lambda}_p\mu^\1_{\pa 1}
        +\hat{\lambda}\Si\20j\epsilon_{pj},\label{eq:dotL31p}
\end{eqnarray}\endnumparts

\begin{eqnarray}
    \var{H}{B^\A_0}&=D_1\Si\A01 +D_i\Si\A0i.\label{eq:dotP}
\end{eqnarray}

We now propagate the primary constraints \eref{2.2:prim} using
$\dot{Z}=\left\{Z,H\right\}$ and obtain
\begin{eqnarray}
    \dot{C}^i&=\mu^\0_{\pa 0}\dot{\tilde{\Sigma}}_\2^{\pa 0i},\label{eq:dotCi}
\end{eqnarray}
\numparts\begin{eqnarray}
    \dot{\tilde{\Sigma}}_\2^{\pa 01}&=\Si {\2\pa,\,i}1i +A^\3_i\Si\11i
        +2A^\1_{i}\Si\21i +B^\3_0\Si\101+B^\1_0\Si\201,\label{eq:dotS201}\\
    \dot{\tilde{\Sigma}}_\3^{\pa 01}&=\Si{\3\pa,\,i}1i
        -A^\2_i\Si\11i -2A^\1_{i}\Si\31i
        -B^\2_0\Si\101 -B^\1_0\Si\301.  \label{eq:dotS301}
\end{eqnarray}\endnumparts

We must now check which of the above equations are secondary equations and
which define multipliers. We first see that~\eref{eq:dotk1p} defines the
multipliers $\lambda^1_p=-(R^\2_{\pa pj}
\Si\21j+R^\3_{pj}\Si\31j)/\Si\101$. Equation \eref{eq:dotp00} then
determines $\hat{\lambda}$ which is given by
$\hat{\lambda}\approx-R^\1_{\pa 23}/\Si\101$ (where the symbol
$\approx$ indicates weak equality in which we ignore terms that vanish by
virtue of the equations of motion). If this is substituted
into~\eref{eq:dotp11} then it becomes weakly zero.  Also, after
substituting $\hat{\lambda}$ into equation~\eref{eq:dotp10}, the multiplier
equation $\lambda_i\Si\21i\approx0$ is obtained. We use~\eref{eq:dotL11p}
to define the multipliers $\lambda^0_p=-R^\1_{\pa 1p}$,
and~\eref{eq:dotL31p} to define $\mu^\1_{\pa 1}\tilde{\lambda}_p= R^\3_{\pa
1p}-R^\3_{\pa ip}s^i_{\pa 1}+R^\1_{\pa
ip}\Si\20i/\Si\101$. Equations~\eref{eq:dotCi} define the cyclic variables
$\Si\21i$, while the final equations~\eref{eq:dotS201}
and~\eref{eq:dotS301} define $B^\2_0$ and $B^\3_0$.  This leaves us with
eight secondary constraints \eref{eq:dotp01}, \eref{eq:dotL21p},
\eref{eq:dotk0p}, \eref{eq:dotP}, which can be written
\numparts
\begin{eqnarray} 
\var{H}{\mu^\1_0}&\ap\Si\30p\left( R^\3_{\pa
1p}\Si\101 +R^\3_{\pa ip}\Si\10i +R^\1_{\pa
ip}\Si\20i\right),\label{eq:secondp}\\ 
\var{H}{\Si\21p}&\ap R^\2_{\pa 1p}\Si\101 +R^\2_{\pa ip}\Si\10i 
+R^\1_{\pa ip}\Si\30i,\label{eq:secondL}\\
\var{H}{s_0^p}&\ap-R^\A_{\pa pj}\Si\A0j 
+R^\A_{\pa 1p}\Sigma_A^{\pa 01},\label{eq:secondk}\\ 
\var{H}{B^\A_0}&\ap D_1\Si\A01
+D_i\Si\A0i.\label{eq:secondP}
\end{eqnarray}
\endnumparts

Therefore at this point we have a phase space of 18 variables, with 4
primary constraints \eref{2.2:prim} and 8 secondary constraints 
\eref{eq:secondp}--\eref{eq:secondP}. We now propagate the
secondary constraints to check for any tertiary constraints. We will
show in the next section that \eref{eq:secondp}, \eref{eq:secondL} and
\eref{eq:secondk} give five of the Einstein equations and
are therefore automatically preserved by the Bianchi identities. When
we propagate \eref{eq:secondP} we find that one component is identically 
zero on the reduced phase space, whereas the other two components
define the multipliers $\tau^\2$ and $\tau^\3$. No further constraints are
therefore obtained by propagating the secondary constraints.

Now that we have obtained all the constraints we obtain
the evolution equations by making variations with respect to 
the canonical variables. This gives
\numparts\begin{eqnarray}
\fl    \dot{A}^\1_p&=D_pB^\1_0+R^\1_{\pa ip}s^i_{\pa 0}-R^\2_{\pa pj}\Si\21j
        \left(\Si\101\right)^{-1},\label{eq:eomA1i}\\
\fl    \dot{A}^\2_p&=D_pB^\2_0+R^\2_{\pa ip}s^i_{\pa 0}+\mu^\0_{\pa 0}\lambda_p
        -R^\1_{\pa pj}\Si\31j\left(\Si\101\right)^{-1},\label{eq:eomA2i}\\
\fl    \dot{A}^\3_p&=D_pB^\3_0+R^\3_{\pa ip}s^i_{\pa 0} -R^\1_{\pa pj}\Si\21j
        \left(\Si\101\right)^{-1},\label{eq:eomA3i}
\end{eqnarray}\label{eq:eomA}\endnumparts
\numparts\begin{eqnarray}
\fl    \dot{B}^\1_1&=D_1 B^\1_0 +\mu\hat{\lambda} +\lambda^a_is^i_{\pa a},
    \label{eq:eomb11}\\
\fl    \dot{B}^\2_1&=D_1 B^\2_0+\tau^\2,\\
\fl    \dot{B}^\3_1&=D_1 B^\3_0+\tau^\3,
\end{eqnarray}\endnumparts
\numparts\begin{eqnarray}
\fl    \dot{\tilde{\Sigma}}_\1^{\pa 0i}&=
        2D_j\left(\Si\1a{[i}s_{\pa a}^{j]}\right)
        -D_1(\Si\11i)+\epsilon^{ij}(\mu-s\Si\101)_{,j}
        +2\eta^\C_{\pa \B\1}B^\B_0\Si\C0i, \label{eq:dotSi10i}\\
\fl    \dot{\tilde{\Sigma}}_\2^{\pa 0i}&=
        2D_j\left(\Si\2a{[i}s_{\pa a}^{j]}\right)
        -D_1(\Si\21i)+\epsilon^{ij}A^\3_j(\mu
        -s\Si\101)+2\eta^\C_{\pa \B\2}B^\B_0\Si\C0i\\
\fl    \dot{\tilde{\Sigma}}_\3^{\pa 0i}&=
        2D_j\left(\Si\3a{[i}s_{\pa a}^{j]}\right)
        -D_1(\Si\31i)+\epsilon^{ij}A^\2_j(\mu
        -s\Si\101)+2\eta^\C_{\pa \B\3}B^\B_0\Si\C0i, \label{eq:dotSi30i}
\end{eqnarray}\label{eq:dotSiA0i}\endnumparts
\begin{eqnarray}
\fl    \dot{\tilde{\Sigma}}_\A^{\pa 01}&=D_i\Si\A1i
        +2\eta^\C_{\pa \B\A}B^\B_0\Si\C01.\label{eq:dotSiA01}
\end{eqnarray}

\section{Einstein equations}
We now show that the equations which we have obtained so far contain
the ten Einstein equations. In order to do this we
first represent the Einstein equations in terms of the variables used
in the Hamiltonian description. 
\numparts\begin{eqnarray}
\Si\101G^\0_{\pa \0}&\approx 2uv\left(R^\2_{\pa 1j}\Si\101
        +R^\2_{\pa ij}\Si\10i +R^\1_{\pa ij}\Si\30i\right)\Si\21j,\\
    \Si\101G^\0_{\pa \1}&\approx-2(u^1_{\pa \1})^2v\left(
        R^\3_{\pa 1j}\Si\101 +R^\3_{\pa ij}\Si\10i\right)\Si\30j,\\
    \Si\101G^\0_{\pa \2}&\approx -2uv\left(
        R^\1_{\pa 1j}\Si\101 +R^\3_{\pa ij}\Si\30i
        +R^\1_{\pa ij}\Si\10i\right)\Si\21j,\\
    \Si\101G^\0_{\pa \3}&\approx-2(u^1_{\pa \1})^2v
        \left(R^\1_{\pa 1j}\Si\101
        +R^\1_{\pa ij}\Si\10i\right)\Si\30j,\\
    \Si\101G^\2_{\pa \3}&\approx-2(u^1_{\pa \1})^2v
        \left(R^\2_{\pa 1j}\Si\101
        +R^\2_{\pa ij}\Si\10i\right)\Si\30j,\\
    \Si\101G^\1_{\pa \0}&\approx-2(u^0_{\pa \0})^2v
        \left(R^\2_{\pa 0j}\Si\101
        -R^\2_{\pa ij}\Si\11i\right)\Si\21j,\\
    \Si\101G^\1_{\pa \2}&\approx-2(u^0_{\pa \0})^2v
        \left(R^\1_{\pa 0j}\Si\101
        -R^\1_{\pa ij}\Si\11i\right)\Si\21j,\\
    \Si\101G^\1_{\pa \3}&\approx -2uv
        \left(R^\1_{\pa 0j}\Si\101-R^\1_{\pa ij}\Si \11i
        -R^\2_{\pa ij}\Si \21i\right)\Si\30j,\\
    \Si\101G^\3_{\pa \2}&\approx -2uv
        \left(R^\3_{\pa 0j}\Si\101
        -R^\3_{\pa ij}\Si\11i\right)\Si\21j,\\
    \Si\101G^\3_{\pa \3}&\approx 2uv\Big[\left(
        R^\1_{\pa 0i}\Si\101+R^\1_{\pa ij}\Si\11j
        +R^\2_{\pa ij}\Si \21j\right)\Si\10i\nonumber\\
    &\qquad \pa\pa +\left(R^\2_{\pa 1i}\Si \21i
        -R^\1_{\pa 1i}\Si \11i
        +R^\1_{\pa 01}\Si \101\right)\Si\101\Big].
\end{eqnarray}\label{2.2:eineq}\endnumparts
We now see that the first five equations are
determined by the secondary constraints as follows
\numparts\begin{eqnarray}
    \Si\101G^\0_{\pa \0}& \approx
        2uv\Si\21i\var{H}{\Si \21i}=0, \label{G00}\\
    \Si\101G^\0_{\pa \1}& \approx-2(u^1_{\pa\1})^2v\var{H}{\mu^\1_0}=0, \\
    \Si\101G^\0_{\pa \2}& \approx
        -2uv\Si\21i\var{H}{s^i_{\pa 0}}=0,\\
    \Si\101G^\0_{\pa \3}&=
        2(u^1_{\pa \1})^2v\Si\30i\var{H}{s^i_{\pa 0}}=0,\\
    \Si\101G^\2_{\pa \3}& \approx
        2(u^1_{\pa \1})^2v\Si\30i\var{H}{\Si \21i}=0. \label{G23}
\end{eqnarray}\endnumparts
Note that $\Si\101=\nu\neq 0$ so that these do indeed imply the vacuum
Einstein equations.

We now show that the equations of motion 
\eref{eq:eomA1i}--\eref{eq:eomb11} 
express the remaining Einstein equations. Writing 
equation \eref{eq:eomA1i} in the form
\begin{equation}
    -\dot{A}^\1_p+D_pB^\1_0+R^\1_{\pa ip}s^i_0-R^\2_{\pa pj}\Si\21j
        \left(\Si\101\right)^{-1}=0
\end{equation}
and using the definition of $R^\1_{\pa 0i}$ and the constraints $C^i_0$, we
find  
\begin{equation}
-R^\1_{\pa 0p}\Si\101+R^\1_{\pa ip}\Si\11i
        -R^\2_{\pa pj}\Si\21j \approx 0,\label{eq:eomad1}
\end{equation}
which implies that $G^\1_{\pa \2}\ap0$ and $G^\1_{\pa \3}\ap0$.
In a similar way we rewrite the remaining equations \eref{eq:eomA2i},
\eref{eq:eomA3i} and \eref{eq:eomb11} to obtain
\begin{eqnarray}
    \left(-R^\2_{\pa 0p}+R^\2_{\pa ip}\Si\11i\right)
        \Si\21p\approx0,\label{eq:eomad2}\\
    -R^\3_{\pa 0p}\Si\101+R^\3_{\pa ip}\Si\11i
        -R^\1_{\pa pi}\Si\21i\approx0,\label{eq:eomad3}\\
    R^\1_{\pa 01}\Si\101-R^\1_{\pa 1i}\Si\11i
    +R^\2_{\pa 1i}\Si\21i\approx0.\label{eq:eomad4}
\end{eqnarray}\endnumparts
Equation \eref{eq:eomad2} gives $G^\1_{\pa \0}\ap0$, 
whilst \eref{eq:eomad3} gives $G^\3_{\pa \2}\ap0$. The final
Einstein equation $G^\3_{\pa \3}\ap0$ follows from
\eref{eq:eomad1} and \eref{eq:eomad4}. We have therefore shown that the
constraint equations and evolution equations imply the Einstein
equations.

\section{Structure equations}
From the self-dual Lagrangian approach one obtains not only the Einstein
equations but also the structure equations. These are derived through the
variation of the connection variables and when written in terms of the
$SO(3)$ basis give the structure equations,
$dS^\A+2\eta^{\A}_{\pa\B\C}\Gamma^\B \wedge S^\C =0$. When this is
expressed in terms of the sigma variables we obtain the equations $D_\alpha
\Si \A\gamma\alpha=0$ and we should expect to obtain these equations as
well as the Einstein equations from our Hamiltonian analysis. \bigs

We would normally expect the structure equations to come from the equations
of motion, but this is not completely true in this case. The equations of
motion \eref{eq:dotSi10i}-\eref{eq:dotSi30i} and \eref{eq:dotSiA01} 
can be rewritten as
$-D_\alpha \Si \A\alpha i=0$ and $D_\alpha \Si \A1\alpha=0$
respectively. The remaining structure equations are not found in the
equations of motion but in the constraint equation \eref{eq:dotP} which can
be rewritten as $D_\alpha \Si \A\gamma\alpha$; this is a result of using
the shortcut method. Combining these equations we obtain $D_\alpha \Si
\A\gamma\alpha=0$ which when written in terms of $S^\A$ gives us the
structure equations.

\section{First Class constraints}

The constraints obtained so far are not necessarily first class.  We need
to take linear combinations of the four primary and eight secondary
constraints to construct a first class algebra. It is possible to do this
by following the Dirac-Bergmann algorithm but in practice it is easier to
use geometric insight to construct the appropriate variables. For example
in the 2+2 formalism we would expect that two of the first class
constraints $\psi_p$ would generate diffeomorphisms in the 2-surface
$\{S\}$. These will come from the shift terms so we start by considering
the secondary constraints that arise from the variation of the multipliers
$s^p_{\pa 0}$.  By calculating the Poisson brackets with the connection we find
we need to adapt them by the addition of the constraint \eref{eq:dotP},
multiplied with the canonical variables $A^\A_p$. This gives the constraint
\begin{eqnarray}
    \psi_p:&=R^\A_{\pa ip}\Si\A0i +R^\A_{\pa 1p}\Si\A01
        +A^\A_p\left(D_1\Si\A01+D_i\Si\A0i\right) \nonumber \\
    &=B^\A_{1,p}\Si \A01 +A^\A_{i,p}\Si \A0i -(A^\A_p\Si \A01)_{,1}
        -(A^\A_p\Si \A0j)_{,j}=0. \label{2.2:psi_p}
\end{eqnarray}
Another first class constraint $\psi_1$ should correspond to the
re-parameterisation of the null generators of $\Sigma_\0$. In the 2+2
formalism this will come from the lapse of $\{S\}$ in $\Sigma_\0$, 
so we start by considering the constraint generated by $\mu^\1_{\pa 0}$.
Then, to obtain the first class constraint we adapt it in a similar
manner to the previous constraint to obtain
\begin{eqnarray}
    \psi_1:&=R^\A_{\pa i1}\Si\A0i+B^\A_1\left(D_1\Si\A01
        +D_i\Si\A0i\right) \nonumber\\
    &=B^\A_{1,1}\Si \A01 +A^\A_{i,1}\Si \A0i -(B^\A_1\Si \A01)_{,1}
        -(B^\A_1\Si \A0j)_{,j}=0. \label{2.2:psi_1}
\end{eqnarray}
The final first class constraint is a Gauss type equation obtained from 
$\dot{B}^\1_0$,
\begin{eqnarray}
    \m G_\1:=D_1\Si\101+D_i\Si\10i&=0\label{2.2:f4}.
\end{eqnarray}

Based on the usual 3+1 timelike decomposition of the action one might
expect a further first class scalar Hamiltonian constraint. However it has
been observed in a number of studies \cite{torre}, 
\cite{jg&dr:92}, \cite{goldsot} that in a null formulation of Einstein's
equations the Hamiltonian constraint is second class. Geometrically this is
because a null surface is special since there are no compact infinitesimal
mappings from one null surface to another null surface \cite{goldrob}.
In fact one can show that there are no further first class constraints, so
at the end of this analysis we have a phase space of 18 functions subject
to 4 first class and 8 second class constraints leaving 2 dynamical degrees
of freedom per hypersurface point as is appropriate on a null surface
\cite{sachs}, \cite{goldberg}, \cite{goldsot}, \cite{rp:80}.   

We now wish to consider the geometric interpretation of the first
class constraints. In order to do this we calculate the infinitesimal
transformations of the canonical variables generated by the
constraints. 
We start by considering $\psi_i$. Let $\tilde F$ be a vector field with
components tangent to $\{S\}$ and define a smeared version of the $\psi_i$
constraint by
\begin{equation}
\tilde\Psi(\tilde F)=\int \tilde F^i\psi_i \ud^3x.
\end{equation}
Then taking the commutator of this with the connection gives
\numparts\begin{eqnarray}
    \delta A^\A_{\pa i}&=\left\{A^\A_i,\tilde\Psi(\tilde F)\right\}=
        \m L_{\tilde F}A^\A_{\pa i},\\
    \delta B^\A_{\pa 1}&=\left\{B^\A_1,\tilde\Psi(\tilde F)\right\}=
        \m L_{\tilde F}B^\A_{\pa 1},
\end{eqnarray}\label{2.2:infinit_i}\endnumparts
(where $\m L$ denotes the Lie derivative in $\Sigma_\0$) 
which shows that $\psi_i$ generates diffeomorphisms within the two surface
$\{S\}$. 

We next consider $\psi_1$. This time we let $\hat F$ be a vector field with
components tangent to the null generators of $\Sigma_\0$ and defined a
smeared version of the $\psi_1$ constraint by
\begin{equation}
\hat\Psi(\hat F)=\int \hat F^1\psi_1 \ud^3x.
\end{equation}
Taking the commutator of this constraint with the connection then gives 
\numparts\begin{eqnarray}
    \delta A^\A_{\pa i}&=\left\{A^\A_i,\hat\Psi(\hat F)\right\}=
        \m L_{\hat F}A^\A_{\pa i},\\
    \delta B^\A_{\pa 1}&=\left\{B^\A_0,\hat\Psi_1(\hat F)\right\}=
        \m L_{\hat F}B^\A_{\pa 1},
\end{eqnarray}\label{2.2:infinit_1}\endnumparts
which shows that $\psi_1$ generates diffeomorphisms along the null
generators of $\Sigma_\0$.

Finally we consider the constraint $\m G_1$. We smear this constraint with
the scalar field $f$ and define
\begin{equation}
G(f)=\int f \m G_1 \ud^3x.
\end{equation}
The commutator of this constraint with the connection is then given 
by ~\eref{2.2:f4} 
\numparts\begin{eqnarray} \delta
A^\A_{\pa i}&=\left\{A^\A_i,G(f)\right\}=-f_{,i}\delta^\A_\1
-2fA^\2_i\delta^\A_\2 +2fA^\3_i\delta^\A_\3, \label{comm2}\\ 
\delta B^\A_{\pa
1}&=\left\{B^\A_1,G(f)\right\}=-f_{,1}\delta^\A_\1
-2fB^\2_1\delta^\A_\2 +2fB^\3_1\delta^\A_\3.
\label{comm3}
\end{eqnarray} \endnumparts

To understand the transformation generated by this constraint we need to
consider the effect of spin and boost transformations.
A complex spin and boost transformation is given by
\begin{equation}\label{2.2:sp}\begin{array}{l}
  \theta^\0\longrightarrow \rho r\theta^\0,\\
  \theta^\1\longrightarrow \rho^{-1} r^{-1}\theta^\1,\\
  \theta^\2\longrightarrow \rho r^{-1}\theta^\2,\\
  \theta^\3\longrightarrow \rho^{-1} r\theta^\3,
\end{array}\end{equation}
where $r=\bar{\rho}$ in the real case.
This transformation induces the following change in the self-dual 2-forms
$S^\A$
\begin{equation}
    S^\A\rightarrow(\Lambda^{-1})^\A_{\pa \B}S^\B
\end{equation}
where 
\begin{equation}(\Lambda)^\A_{\pa \B}=\left[
        \begin{array}{ccc}
         1 & 0 & 0 \\
          0 & r^2 & 0 \\
          0 & 0 & r^{-2} \\
        \end{array}\right].
\label{gauge}
\end{equation}
Note that this only depends upon $r$ and not on $\rho$ which reflects the
fact that $r$ represents the self-dual part and $\rho$ the anti self-dual
part of the spin and boost freedom.

Under a gauge transformation \eref{gauge}
the connection transforms according to
\begin{equation}
    \Gamma^\A\longrightarrow\eta_\B^{\pa \A\C}(\Lambda^{-1})^\B_{\pa \D}
        \ud(\Lambda)^\D_{\pa \C} +\eta_\B^{\pa \A\C}\eta^\F_{\pa \D\E}
        (\Lambda^{-1})^\B_{\pa \F}(\Lambda)^\D_{\pa \C}\Gamma^\E.
\end{equation}
Using this we find the infinitesimal transformations of the connection
variables, $A^\A_i$ and $B^\A_a$, are given by
\numparts
\begin{eqnarray}\label{2.2:deltaAB}
          \delta A^\1_i\rightarrow\delta r_{,i}, \\
          \delta B^\1_a\rightarrow\delta r_{,a}, \\
          \delta A^\2_i\rightarrow -2A^\2_i \delta r,\\
          \delta B^\2_a\rightarrow -2B^\2_a \delta r,\\
          \delta A^\3_i\rightarrow 2A^\3_i \delta r,\\
          \delta B^\3_a\rightarrow 2B^\3_a \delta r.
\end{eqnarray}
\endnumparts
Comparing this to \eref{comm2} and \eref{comm3} we see that 
$\m G_\1$ generates the self-dual spin and boost transformations.

We are now in a position to display the structure of the first class
algebra. We do this by calculating the Poisson brackets of all the smeared
first class constraints with each other. This has the following structure
\numparts
\begin{eqnarray}
\left\{\tilde\Psi(\tilde P),\tilde\Psi(\tilde Q)\right\}=
\tilde\Psi(\m L_{\tilde P}\tilde Q),\\
\left\{\tilde\Psi(\tilde P),\hat\Psi(\hat Q)\right\}=
\hat\Psi(\m L_{\tilde P}\hat Q),\\
\left\{\hat\Psi(\hat P),\hat\Psi(\hat Q)\right\}=
\hat\Psi(\m L_{\hat P}\hat Q),\\
\left\{\tilde\Psi(\tilde P),G(q)\right\}=
G(\m L_{\tilde P}q),\\
\left\{\hat\Psi(\hat P),G(q)\right\}=
G(\m L_{\hat P}q),\\
\left\{G(p),G(q)\right\}=
0.
\end{eqnarray}
\endnumparts
We have chosen to keep $\psi_1$ and $\psi_i$ separate to illustrate the 2+2
structure of the constraint algebra. However they may be combined to give 
$\psi_A$, where $(\psi_A)=(\psi_1, \psi_2, \psi_3)$. This may be smeared
with a general vector field $F$ on $\Sigma_0$ to give 
\begin{equation}
\Psi(F)=\int F^A\psi_A \ud^3x.
\end{equation}
The constraint algebra then has the more compact form
\numparts
\begin{eqnarray}
\left\{\Psi(P),\Psi(Q)\right\}=
\Psi(\m L_{P} Q),\\
\left\{\Psi(P),G(q)\right\}=
G(\m L_{P}q),\\
\left\{G(p),G(q)\right\}=
0.
\end{eqnarray}
\endnumparts
This algebra has a similar form to that obtained by Goldberg 
\emph{et al.} \cite{jg&dr:92} but is simpler because in our 2+2 case $G$
generates the self-dual spin and boost transformations rather than the more
complicated null rotations which are needed in the null 3+1 setting.

\section{Second class constraints and reality conditions}
In this section we briefly examine the remaining constraints. We start by
looking at the geometric origin of the second class constraints. See
Goldberg and Robinson \cite{goldrob} for a similar analysis. 
The double null slicing condition
together with the use of an adapted frame results in four constraints
\begin{equation}
    \Si \20i=0,\qquad \Si \201=0,\qquad \Si \301=0.\label{sec1}
\end{equation}
The invariance of null directions in the hypersurface under spin and boost
transformations gives a further two second class constraints given by
equation \eref{eq:secondP} with $\A=2,3$, which we denote 
\begin{equation}
{\m G}_\2=0,\qquad {\m G}_\3=0.\label{sec2}
\end{equation}
The remaining two second class constraints are given by
\eref{eq:secondL} with $p=2,3$. However rather than work with these
constraints we use instead the linearly independent combinations 
\begin{equation}
{\m H}_0=\Si\21i\var{H}{\Si \21i}=0,
\end{equation}
and
\begin{equation}
\phi=\Si\30i\var{H}{\Si \21i}=0.
\end{equation}
We see from the analysis of section 5 that according to \eref{G00} 
${\m H}_0$ generates the 
$G^\0_{\pa \0}$ component of the Einstein equations and hence corresponds
to the usual scalar Hamiltonian constraint which as expected is 
second class when using a null evolution \cite{torre}, \cite{aragone}, 
\cite{jg&dr:92}, \cite{goldsot}. We also see from \eref{G23} that $\phi$
corresponds to the Einstein equations in the two surface $S$. 

The Poisson bracket algebra of the second class constraints is quite
complicated and not very illuminating. However the general structure of the
algebra can be seen by defining a vector of second class constraints by
\begin{equation}
C_I=\left({\m G}_2, {\m G}_3,  {\m H}_0, \phi, 
\Si \201, \Si \202, \Si \203, \Si \301 \right)
\end{equation}
for $I=1,\ldots, 8$. The Poisson bracket matrix then has the structure
\begin{equation}
C=\left(\begin{array}{cc}
         Q & R \\
         -\tilde R & 0\\
        \end{array}\right), \label{secondalg}
\end{equation}
where $Q$ and $R$ are $4\times4$ matrices. Note this has a similar form to
that of Goldberg and Robinson \cite{goldrob}.

We now turn to the reality conditions. Since we are using an adapted frame
the relevant conditions are
\begin{equation}
\mu^{\2}_{\pa b}=\bar\mu^{\3}_{\pa b},\qquad
\nu^{\2}_{\pa j}=\bar\nu^{\3}_{\pa j}.
\end{equation}
However rather than use the $\nu^{\i}_{\pa j}$ variables we have chosen to
use instead the mixed terms of the densitised 2-forms. We therefore also
require the reality conditions for the 2-forms which are given by
\begin{equation}
\epsilon_{\ba\bb\bc\bd}\Si A\ba\bb \bar{\tilde{\Sigma}}_\B^{\pa \bc\,\bd}=0.
\end{equation}

\section{Conclusion}
In this paper we have applied a canonical analysis to a double null
description of General Relativity formulated in terms of Ashtekar type
variables. We started from a first order action written in terms of
self-dual two forms and the curvature of a complex $SO(3)$ connection and
used this to obtain a Lagrangian density in terms of our variables.  From
this we calculated the Hamiltonian, on which we performed the canonical
analysis. We obtained four primary constraints and eight
linearly independent secondary constraints.  By taking particular linear
combinations of these twelve constraints, we revealed four first class
constraints. Two of these constraints, $\psi_p$, generate the
diffeomorphisms within the spatial hypersurface $\{S\}$; while one
constraint, $\psi_1$, generates the diffeomorphisms along the null
generators of $\Sigma_\0$. The final first class constraint, is the Gauss
constraint which generates the self-dual spin and boost transformations.
Unlike the case in the standard 3+1 description, the constraint algebra
forms a Lie algebra. This results from  using a null formulation in
which the Hamiltonian constraint (which causes all the difficulties) is no
longer a first class constraint, but because of the null formulation, is now 
second class.

The next step of the canonical quantisation process would be to pass to a
reduced phase space which represents the true degrees of freedom of the
theory. This involves restricting to the phase space where the second class
constraints are satisfied but replacing the Poisson brackets by Dirac
brackets \cite{Dirac} (see also Isenberg and Nester \cite{gr&g}). 
These are modified versions of the Poisson brackets such that the Dirac
bracket between any of the second class constraints and any other variable
vanishes identically. Given two functions $F$ and $G$ on the phase space
the Dirac bracket is given by
\begin{equation}
\{F,G\}_D=\{F,G\}-\sum_{J,K}\{F,C_J\}C^{-1 JK}\{C_K,G\},
\end{equation}
where $C^{-1}$ is the inverse of the matrix given by \eref{secondalg}.

A similar result is obtained by using instead the `starred variables' of 
Bergmann and Komar \cite{bergkomar} which are constructed so as to have 
vanishing Poisson bracket with all the second class constraints. An
alternative approach would be to fix the gauge in some appropriate way and
explicitly solve for the constraints to obtain a Hamiltonian in terms of
the independent degrees of freedom. This procedure has been carried out for
$D$-dimensional gravity in the light-cone gauge by Goroff and Schwartz
\cite{goroffschwartz},  (see also \cite{aragone})
but the complicated nature of the second class
constraints in our case makes it unlikely that one can do this at all
simply in the present formalism.

In the formulation of General Relativity used here some of the
variables, $\mu^\a_{\pa b}$, contain an anti self-dual part. 
Because the action is written in terms of self-dual two forms this could
lead to complications. In the canonical analysis given above we were able 
to avoid this difficulty because these variables 
were also cyclic, so we we could treat
them as multipliers rather than canonical variables. However this problem
is likely to be more serious when one moves on to the next step of the
process. For this reason it is desirable to eliminate the frame variables
entirely (as has already been done with the $\nu^\i_{\pa j}$ terms) and work
exclusively with the components of $\Sigma_\A$ which are manifestly
self-dual. This approach has been developed in a subsequent paper \cite{dvl}.

\end{document}